\documentclass[preprint,showpacs,preprintnumbers,amsmath,amssymb]{revtex4}
\usepackage{graphicx}
 \usepackage{longtable}
 \usepackage{dcolumn}
 \usepackage{bm}
\usepackage{color}

\begin{document}

\title{Existence of hyperons in the pulsar PSRJ1614-2230 }
\author{A. Sulaksono$^1$ }
\author{B. K. Agrawal$^2$, }
\affiliation{
$^1$ Departemen Fisika, FMIPA, Universitas Indonesia, Depok, 16424,
Indonesia.  \\
$^2$Saha Institute of Nuclear Physics, Kolkata - 700064, India.
}
\begin{abstract}

The possibility of existence of hyperons in the recently measured
$2M_\odot$ pulsar PSRJ1614-2230 is explored using a diverse set
of nuclear equations of state calculated within  the relativistic
mean-field models. Our results  indicate that the nuclear equations of
state compatible with heavy-ion data allow the hyperons to exist in the
PSRJ1614-2230 only for significantly larger values for the meson-hyperon
coupling strengths.  The maximum mass configurations for these cases
contain sizable hyperon fractions  ($\sim 60\%$) and yet masquared
their counterpart composed of only nucleonic matter.

 \end{abstract}
\preprint{APS/123-QED}
   \maketitle

\section{Introduction}
\label{sec:intro}

The latest measurement of the Shapiro delay  for the millisecond
pulsar PSRJ1614-2230 provides reliable lower bound on the maximum
mass to be $1.97\pm 0.04M_\odot$ \cite{Demorest10}.   This measurement
rules out all the equations of state (EOSs) yielding the maximum
mass less than that of the PSRJ1614-2230.  Of course, the EOSs
for the nucleonic matter can readily yield the compact stars with
masses $\sim 2M_\odot$. The EOSs with hadron-quark phase transition
are also compatible with the mass measurement of the PSRJ1614-2230,
provided, the quarks are assumed to be strongly interacting and are
in colour superconducting phase \cite{Ozel10a,Weissenborn11a}.
However, at large, the maximum mass of the compact stars
are found to be well below  $2M_\odot$ when  the non-nucleonic
degrees of freedom like hyperons and kaon condensates are considered
\cite{Glendenning91,Glendenning98,Lackey06,Schulze06,Lattimer07,Bednarek07}.
One might thus infer in the backdrop of previous calculations that
the existence of hyperons and kaon condensates are unlikely in the
PSRJ1614-2230.

Recently, studies involving role of hyperons on
the maximum mass of the compact stars are revisited
\cite{Stone10,Bednarek11,Trumper11,Weissenborn11,Logoteta12}.  It is
shown that the EOS for the hyperonic matter at higher densities
can be stiffened in several ways within the relativistic mean-field
(RMF) models.  The maximum mass for these EOSs are above $2M_\odot$
indicating the existence of hyperons in the PSRJ1614-2230.  The EOS of
the hyperonic matter are stiffened by including the vector-isoscalar
strange $\phi$ mesons within the RMF model, in addition to the usual
scalar-isoscalar $\sigma$, vector-isoscalar $\omega$ and vector-isovector
$\rho$ mesons \cite{Weissenborn11}.  The $\phi$ mesons stiffens the
EOS   at densities far beyond the nuclear saturation density. Further,
the increase of about $ 0.2M_\odot$ in the  maximum mass of the compact
stars is achieved by varying the hyperon-nucleon potential depths
for the $\Sigma$ and $\Xi$ hyperons which are not very well known.
It has been also demonstrated \cite{Weissenborn11b} that maximum mass
of the compact stars containing hyperons are well above $2 M_\odot$,
provided, the $\omega$ mesons are coupled to hyperons as strongly as
to the nucleons. Such strong meson-hyperon couplings stemmed from
the breaking of SU(6) symmetry.  Another way to stiffen the EOS is
to include the non-linear terms for the $\phi$-mesons as suggested in
Ref. \cite{Bednarek11}.  It may be pointed out that the nuclear EOSs
employed in Ref.  \cite{Weissenborn11,Weissenborn11b} are generally
quite stiff in comparison to  those extracted from the heavy-ion data
\cite{Aichelin85,Fuchs01,Fuchs08,Danielewicz02}.

An alternative scenario has also emerged in which compact stars
containing hyperons can have masses $\sim 2M_\odot$. In this scenario,
one assumes the presence of hypothetical weakly interacting light bosons
(WILBs) at densities several times of the nuclear saturation density.  The
role of WILBs on the EOS or the internal structure of the compact stars
crucially depends on the choice of the characteristic scale which is the
square of the ratio of the coupling strength to the masses of WILBs.  The
values of the characteristic scale of the WILBs are poorly constrained
at present by the laboratory data and the compact star observable
\cite{Lamoreaux97,Decca05,Mostepanenko08,Krivoruchenko09}. Consequently,
the  nuclear EOSs compatible with the heavy-ion data allow the
hyperons to exist in the compact stars with mass $\sim 2M_\odot$
\cite{Sulaksono11}. The required stiffness of the EOS at very high
density is achieved by adjusting the characteristic scale of the WILBs.

In the present work we explore the possibility of existence
of hyperons in the PSRJ1614-2230 using a diverse set of nuclear EOSs
calculated within the RMF models.  In particular,  we examine whether the
nuclear EOSs compatible with the heavy-ion data can allow the
hyperons to exist in the PSRJ1614-2230 without recourse to the WILBs. This
makes our present investigation quite different than the ones performed
earlier. Most of the previous investigations either employed the nuclear
EOSs which are stiffer in comparison to the heavy-ion data or included
the contributions of the  WILBs.

The paper is organized as follows. In Sec. II we describe the theoretical
framework based on the RMF model.  In Sec. III,  we present our results
for the EOSs for the matter without and with hyperons.  In Sec. IV we
present our results for the bulk properties of the static compact stars.
Finally, we state our conclusions in Sec. V.

\section{ Theoretical frame work }
\label{sec:framework}

We use standard and extended versions of the RMF models to compute the
bulk properties of the compact stars. The standard RMF model includes
the contributions from the non-linear self-interaction for the $\sigma$
meson  and extended RMF model includes the contributions from the self
and/or cross interaction terms for the $\sigma$, $\omega$ and $\rho$
meson. We will discuss both versions of the RMF models.  The derivations
of the effective Lagrangian density and corresponding energy density
functionals  for the extended RMF model are well documented in
Refs.  \cite{Furnstahl96,Serot97,Furnstahl97}.  In RMF models, baryons
interact through the exchange of scalar ($\sigma$), vector ($\omega$
and $\rho$) as well as two additional hidden-strangeness ($\sigma$* and
$\phi$) mesons. The baryons considered in this work are nucleon ($N$)
and hyperons ($\Lambda,\Sigma,\Xi$).  The total Lagrangian density of
RMF model including hyperons plus leptons (l) can be written as,

\begin{equation}
{\mathcal{L}} = {\mathcal{L}}^{\rm free}_{B} + {\mathcal{L}}^{\rm free}_{M} + 
  {\mathcal{L}}^{\rm lin}_{BM} + {\mathcal{L}}^{\rm nonlin}+{\mathcal{L}}^{\rm free}_{l},   
\end{equation}

where the free baryons Lagrangian density is,
\begin{equation}
{\mathcal{L}}^{\rm free}_{B}=\sum_{B=N,\Lambda,\Sigma,\Xi}\overline{\Psi}_B[i\gamma^{\mu}\partial_{\mu}-M_B]\Psi_B,
\end{equation}
Here,  $\Psi_B$ is baryons field and the sum is taken over $N$, $\Lambda$, $\Sigma$,and $\Xi$ baryons. The Lagrangian density for free mesons involved is, 
\begin{eqnarray}
{\mathcal{L}}^{\rm free}_{M}&=&\frac{1}{2}(\partial_{\mu}\sigma\partial^{\mu}\sigma-m_{\sigma}^2\sigma^2)+\frac{1}{2}(\partial_{\mu}\sigma^*\partial^{\mu}\sigma^*-m_{\sigma^*}^2\sigma^{*2})\nonumber\\ &-&\frac{1}{4}\omega_{\mu\nu}\omega^{\mu\nu}+\frac{1}{2}m_{\omega}^2\omega_{\mu}\omega^{\mu}-\frac{1}{4}\phi_{\mu\nu}\phi^{\mu\nu}+\frac{1}{2}m_{\phi}^2\phi_{\mu}\phi^{\mu}\nonumber\\ &-&\frac{1}{4}\mathbf{\rho}_{\mu\nu}\mathbf{\rho}^{\mu\nu}+\frac{1}{2}m_{\rho}^2\mathbf{\rho}_{\mu}\mathbf{\rho}^{\mu}.
\end{eqnarray}
The $\omega^{\mu\nu}$, $\phi^{\mu\nu}$ and $\mathbf{\rho}^{\mu\nu}$ are field tensors
corresponding to the $\omega$, $\phi$ and $\rho$ mesons field, and can be defined as
$\omega^{\mu\nu}=\partial^{\mu}\omega^{\nu}-\partial^{\nu}\omega^{\mu}$, $\phi^{\mu\nu}=\partial^{\mu}\phi^{\nu}-\partial^{\nu}\phi^{\mu}$
and $\mathbf{\rho}^{\mu\nu}=\partial^{\mu}\mathbf{\rho}^{\nu}-
\partial^{\nu}\mathbf{\rho}^{\mu}$. The Lagrangian ${\mathcal{L}}^{\rm lin}_{BM}$ describing the interactions of the baryons through the mesons is, 
\begin{equation}                   
{\mathcal{L}}^{\rm lin}_{BM}= \sum_{B=N,\Lambda,\Sigma,\Xi}\overline{\Psi}_B[g_{\sigma B} \sigma + g_{\sigma^* B} \sigma^* -\gamma_\mu g_{\omega B} \omega^\mu-\frac{1}{2}\gamma_\mu g_{\rho B}\mathbf{\tau_B}\cdot \mathbf{\rho} ^\mu -\gamma_\mu g_{\phi B}\phi ^\mu ]\Psi_B,
\label{eq:Llin}
\end{equation}
where $\tau_B$ are the baryons isospin matrices. The Lagrangian describing
nucleons self interactions for $\sigma$, $\omega$, and $\rho$ mesons can be
written as,
\begin{eqnarray}
 {\mathcal{L}}^{\rm nonlin} &=& - \frac{\kappa_3 g_{\sigma N} m_{\sigma}^2}{6 m_{ N} } \sigma^{3}
                   - \frac{\kappa_4 g_{\sigma N}^2 m_{\sigma}^2}{24 m_{ N}^2 } \sigma^{4}+\frac{\zeta_0 g_{\omega N}^2}{24} 
                   {(\omega_{\mu}  \omega^{\mu})}^2\nonumber\\
&+& \frac{\eta_1 g_{\sigma N} m_{\omega}^2}{2 m_{ N} } \sigma \omega_{\mu}  \omega^{\mu}+\frac{\eta_2 g_{\sigma N}^2 m_{\omega}^2}{4 m_{ N}^2 }\sigma^{2} \omega_{\mu}  \omega^{\mu} 
\nonumber\\&+&\frac{\eta_{\rho} g_{\sigma N} m_{\rho}^2}{2 m_{ B} } \sigma
\mathbf{\rho}_{\mu} \cdot \mathbf{\rho}^{\mu} +\frac{\eta_{1\rho}g_{\sigma
N}^2m_{\rho }^{2}}{4m_N^2} \sigma^2\mathbf{\rho}_{\mu} \cdot \mathbf{\rho}^{\mu} +\frac{\eta_{2 \rho} g_{\omega N}^2 m_{\rho}^2}{4 m_{ N}^2 } \omega_{\mu}  \omega^{\mu} \mathbf{\rho}_{\mu} \cdot \mathbf{\rho}^{\mu}.
\label{eq:Lnlin}
\end{eqnarray}
 While the free leptons Lagrangian density is,
\begin{equation}
{\mathcal{L}}^{\rm free}_{l}=\sum_{l=e^-, \mu^-}\overline{\Psi}_l[i\gamma^{\mu}\partial_{\mu}-M_l]\Psi_l.
\end{equation}
here $\Psi_l$ is leptons (electron and muon) field.

The meson-nucleon coupling constants and nonlinear parameters of RMF
models are determined by adjusting them to reproduce ground state
properties of finite nuclei and nuclear matter. Here, we employed NL3,
GM1, TM1, G2, BSR2, BSR6, BSR9, BSR13 and BSP parameter sets for
meson-nucleon coupling constants and nonlinear parameters
\cite{Lalazissis97,Glendenning91,Furnstahl97,Dhiman07,Agrawal10}.
These parameterizations of the RMF models are associated with different
form for the Lagrangian density.  They differ in their non-linear part
of the effective Lagrangian (Eq. \ref{eq:Lnlin}) which are summarized in
Table \ref{tab:rmf}. For instance, the Lagrangian density associated with
the parameter sets NL3 and GM1 correspond to the standard RMF model which
includes non-linear term only for the self-interactions of the $\sigma$
meson. Other parameter sets TM1, G2 and BSR correspond to the extended
RMF models which include contributions from the self-interactions of
the $\omega$ mesons and/or cross-interactions between $\sigma, \omega$
and $\rho$ mesons.

The meson-hyperon coupling strengths $g_{mH}$($m = \sigma, \omega,
\rho, \sigma^{*},\phi$ and $H=\Lambda, \Sigma, \Xi$)
in Eq. (\ref{eq:Llin}) can not be determined very well at present.
The properties of compact stars are quite sensitive to the choices
for the values of $g_{\sigma H}$,$g_{\omega H}$ and $g_{\phi H}$. The
variations in the values of $g_{\rho H}$ and $g_{\sigma^* H}$  do
not appreciably alter the properties of compact stars.  The values of
$g_{\rho H}$ and $g_{\sigma^* H}$ are kept fixed in our calculations.
We take $g_{\sigma^*H} = 0$, this choice is consistent with
$\Lambda\Lambda-$hypernuclear data which yields weak $\Lambda\Lambda$
interaction \cite{Gal11}. The values of $g_{\rho H}$ are taken from the
SU(6) quark model.  We make several choices for the values of $g_{\omega
H}$ and $g_{\phi H}$. One of which correspond to those obtained within
the SU(6) quark model.  The SU(6) quark model values for $g_{\omega H}$,
$g_{\rho H}$ and $g_{\phi H}$ are
  \begin{eqnarray}
\frac{1}{3}g_{\omega N}&=&\frac{1}{2}g_{\omega
\Lambda}=\frac{1}{2}g_{\omega \Sigma}=g_{\omega \Xi},\nonumber\\
g_{\rho N}&=&\frac{1}{2}g_{\rho \Sigma}=g_{\rho \Xi},~ ~ ~ ~ ~
~ g_{\rho\Lambda}=0, \nonumber\\ 2 g_{\phi \Lambda}&=&2 g_{\phi
\Sigma}=g_{\phi \Xi}=\frac{2 \sqrt{2}}{3}g_{\omega N},~ ~ ~ ~ ~ ~
g_{\phi  N}=0.  
\label{eq:su6}
\end{eqnarray} 
For given values of $g_{\omega H}$ the coupling strengths $g_{\sigma H}$
are usually obtained from potential depths for hyperons in the symmetric
nuclear matter at the saturation density as,
  \begin{equation} U_{H}^{(N)}(\rho_s)
= -g_{\sigma H}\sigma(\rho_s)+g_{\omega H}\omega(\rho_s).
  \label{eq:uyn} \end{equation}
The values of potential depths $U_H^{(N)}$ chosen are as follows
\cite{Schaffner00},
 \begin{equation} U_{\Lambda}^{(N)}= -28\text{
MeV}, \qquad U_{\Sigma}^{(N)} = +30\text{ MeV} \qquad \text{and} \qquad
U_{\Xi}^{(N)} = -18\text{ MeV}.  \label{eq:depth}
 \end{equation}
For the sake of convenience we define,

\begin{eqnarray} 
X_{\omega H}=\left\{ \begin{array}{cc} \left(
\frac{g_{\omega H}}{g_{\omega N}}\right)& \text{ for $\Lambda$ and $\Sigma$
hyperons}\\ \\ 2\left( \frac{g_{\omega H}}{g_{\omega N}}\right)& \text{ for
$\Xi$  hyperons}, \end{array} \right .  \label{eq:xw} \end{eqnarray}

and,  \begin{eqnarray} 
\label{eq:xphi}
X_{\phi H}=\left\{ \begin{array}{cc} \left(
\frac{g_{\phi H}}{g_{\omega N}}\right)& \text{ for $\Lambda$ and $\Sigma$
hyperons}\\ \\ \frac{1}{2}\left( \frac{g_{\phi H}}{g_{\omega N}}\right)& 
\text{ for $\Xi$  hyperons}. \end{array} \right . \end{eqnarray}

It has been shown earlier in Ref. \cite{Glendenning91} that the values
of $X_{\omega H}\approx 1$, which are larger compared to those for
the  SU(6) model (Eq. \ref{eq:su6}), can yield  heavier
compact stars composed of hyperons without affecting the properties of
the $\Lambda-$hypernuclei.  Recent investigation \cite{Weissenborn11b}
suggests that the values of $X_{\omega H}$ can significantly differ from
those given by the SU(6) quark model.  The breaking of the SU(6) symmetry
yields $X_{\omega \Lambda} = X_{\omega \Sigma} = \frac{1}{2}X_{\omega
\Xi} = 1$ (i.e., $g_{\omega \Lambda} = g_{\omega \Sigma} = g_{\omega \Xi}
=g_{\omega N}$) . In the present work we vary the values of $X_{\omega
H }$ and $X_{\phi H}$ over a wide range to see in what limits   the
compact stars with hyperons can satisfy the constraints on the lower
bound on the maximum mass imposed by the PSRJ1614-2230.

\section{Equations of state}
\label{sec:eos}

We use different versions of the RMF models, as summarized in Table
\ref{tab:rmf}, to compute the EOSs for the $\beta$-equilibrated
nucleonic and hyperonic matters. Our choice for the models are such
that they yield wide variations in the   various quantities associated
with symmetric and asymmetric nuclear matter at the saturation density.
In Table \ref{tab:nmpro} we provide the values of some of the quantities
associated with nuclear matter at the saturation density $\rho_s$, namely,
the binding energy per nucleons $(B/A)$, incompressibility coefficient
for symmetric nuclear matter ($K$), symmetry energy coefficient ($E_{\rm
sym}$), linear density dependence of the symmetry energy coefficient
($L$) and various quantities ($K_{\rm sym}$), ($K_{\rm asy}$) and
($K_{\rm sat2}$). These quantities are evaluated as follows,

  \begin{eqnarray}
E_{\rm sym}(\rho)=\frac{1}{2}\left .\frac{d^2E(\rho,\delta)}{d\delta^2}\right
|_{\delta=0},\\
L=3\rho_s\left .\frac{d E_{\rm sym}(\rho)}{d\rho}\right |_{\rho=\rho_s},\\
K_{\rm sym}=9\rho^2_s\left .\frac{d^2 E_{\rm sym}(\rho)}{d\rho^2}\right
|_{\rho=\rho_s},\\
\end{eqnarray}
 \begin{eqnarray}
\label{eq:k0}
K=9\rho^2_s\left .\frac{d^2 E(\rho,0)}{d\rho^2}\right
|_{\rho=\rho_s},\\
J=27\rho^3_s\left .\frac{d^3 E(\rho,0)}{d\rho^3}\right
|_{\rho=\rho_s},\\
\label{eq:ksat2}
K_{\rm sat,2}=K_{\rm asy}-\frac{J}{K}L,\\
K_{\rm asy}=K_{\rm sym}-6L.
\end{eqnarray}
where, $E(\rho,\delta)$ is the energy per nucleon at a given density
$\rho$ and asymmetry $\delta=(\rho_n - \rho_p)/\rho$.  In Table
\ref{tab:nmpro}, we also list the values of the neutron-skin $\Delta R$
for the $^{208}$Pb nucleus.  The values of $K$, $E_{\rm sym}$ and $L$
vary over a wide range for the selected RMF model.  This immediately
indicates that the behaviour of the EOSs for the different RMF models
considered should be quite different even at the low densities.

\begin{figure}
\resizebox{6.0in}{!}{ \includegraphics[]{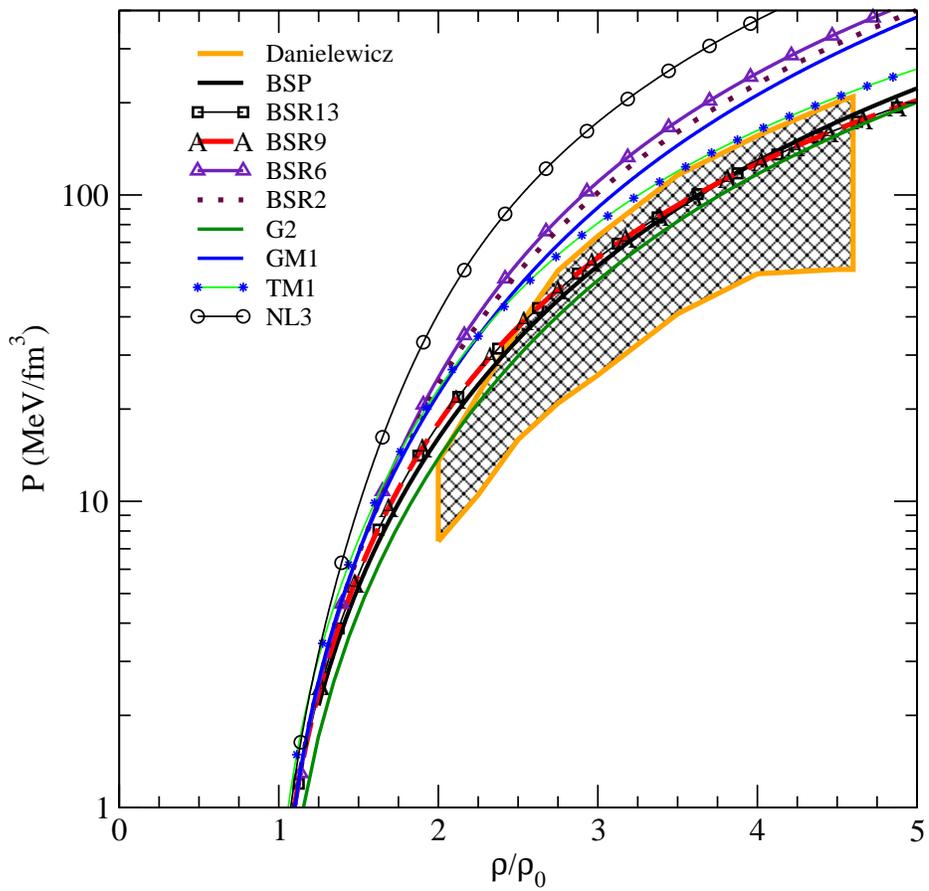}}
\caption{\label{fig:snm} (Color online)
Pressure as a function of nucleon density for the symmetric nuclear matter.
The shaded area represents the EOS extracted from the analysis of Ref.
\cite{Danielewicz02}. The density is scaled by $\rho_0 =0.16$ fm$^{-3}$.}
  \end{figure}

In Fig.  \ref{fig:snm} we plot the EOSs for symmetric nuclear matter
(SNM) in terms of pressure versus nucleon density obtained for various
RMF models.  The bounds on the EOSs as shown by shaded regions are
the ones extracted from heavy-ion collision data \cite{Danielewicz02}.
The EOSs for the NL3, GM1, BSR2 and BSR6  are stiffer in comparison
to those obtained from the heavy-ion data. Whereas, the TM1, G2, BSR9,
BSR13 and BSP are consistent with the EOS from the heavy-ion data. 
The EOS for the BSR2 and BSR6 parameterizations are very much the same
for the reasons as follow.  Both of the parameter sets belong to the same
type of the RMF model as can be seen from Table \ref{tab:rmf}. Further,
the parameters of the BSR2 and BSR6 parameterizations  are obtained by
fit to   exactly same set of experimental data for bulk properties of
finite nuclei, but, for different values of the neutron-skin thickness
in $^{208}$Pb nucleus.  Thus, the BSR2 and BSR6 parameterizations  are
expected to yield different behaviour only for the EOSs of the asymmetric
nuclear matter.  Similar is the case for the EOSs obtained  with the
BSR9 and BSR13 parameterizations.

\begin{figure}
\resizebox{6.0in}{!}{ \includegraphics[]{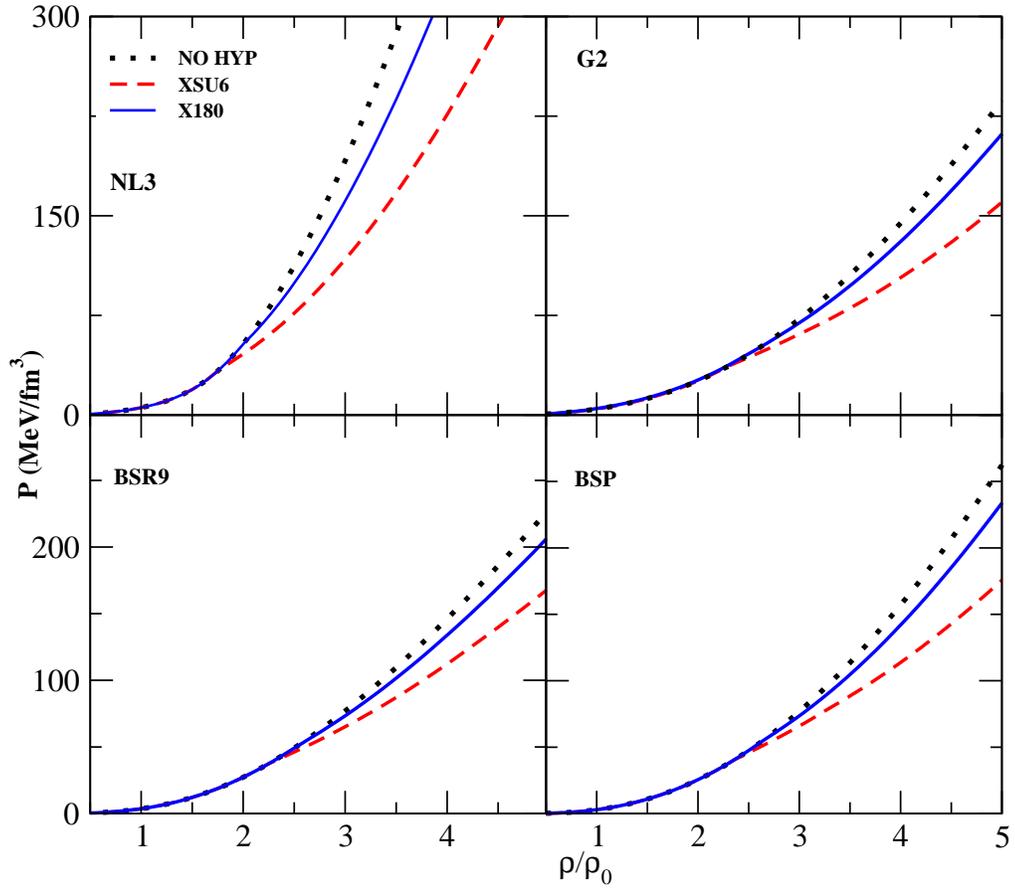}}
\caption{\label{fig:eos_bem} (Color online)
The EOSs in terms of pressure versus density for few RMF models. The
curves labeled 'NO HYP' correspond to the EOSs for the nucleonic
matter. The EOSs for the hyperonic matter for two different choices
for the meson-hyperon coupling strengths are labeled by XSU6 and X180
(see text for detail).
}
  \end{figure}

Our main goal is to investigate whether the nuclear EOSs which are
compatible with the heavy-ion data can allow the hyperons to exist in
the $2M_\odot$ pulsar PSRJ1614-2230.  For this purpose, the EOSs for the
$\beta-$equilibrated hyperonic matter are calculated using different
values for $X_{\omega H}$ and $X_{\phi H}$. We present here the EOSs
only for two different choices for the $X_{\omega H}$ and $X_{\phi
H}$. We refer these choices as XSU6 and X180.  For the case of XSU6,
$X_{\omega H}=\frac{2}{3}$ and $X_{\phi H} = \frac{\sqrt{2}}{3}$ which
correspond to those given by the SU(6) quark model as conventionally
used. The X180 refers to $X_{\omega H} = 1$ and $X_{\phi H} = 0.8$.
The values of $\sigma - H$ coupling strength are determined using
the nucleon-hyperon potential depths as described in Sec. II.
It can be easily verified from Eqs. (\ref{eq:su6},  \ref{eq:xw} and
\ref{eq:xphi}) that the values of $X_{\omega H}$ and $X_{\phi H}$ for
the case of X180 are augmented by a factor of $\sim 1.5$  with respect to their
values for the SU(6) model.  In Fig. \ref{fig:eos_bem}, the EOSs for
the nucleonic and hyperonic matters are compared for a few different
RMF models. The EOSs for the nucleonic matter are labeled as 'NO HYP'.
The EOSs for the hyperonic matter are labeled as XSU6 and X180.  It can
be seen that the EOSs for the hyperonic matter corresponding to X180
are quite close to those obtained for the nucleonic matter.
It should be pointed out that the threshold density is lowest for the
$\Lambda$ hyperons for all the RMF models  considered irrespective of the
choice of the meson-hyperon coupling strengths. Further, the augmented
meson-hyperon coupling strengths results in marginal increase in the
values of the threshold densities for the $\Lambda$ hyperons. For
instance, the threshold densities for $\Lambda$ hyperons for the
NL3(BSP) parameterizations for the XSU6 and X180 cases are 0.28(0.38)
and 0.32(0.42) $\mathrm{fm}^{-3}$,respectively. Thus, the augmented
meson-hyperon coupling strengths might increases the maximum mass of
the compact stars  without significantly affecting its hyperonic contents.

\begin{figure}
\resizebox{5.0in}{!}{ \includegraphics[]{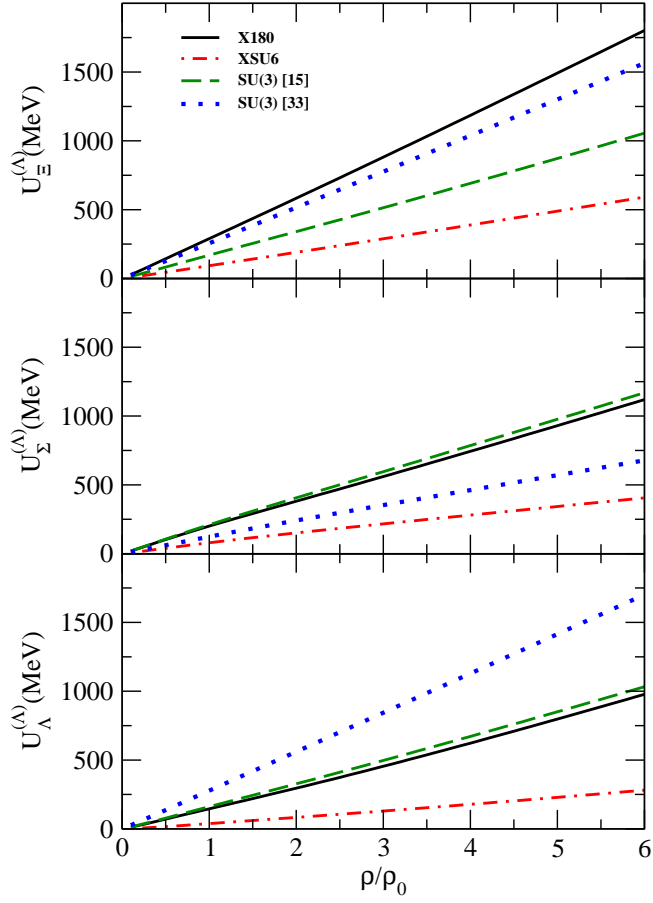}}
\caption{\label{fig:ulh} (Color online)
The $\Lambda-$H potentials $U^{(\Lambda)}_H$ as a function of density
for the meson-hyperon coupling strengths corresponding to XSU6 and X180.
For the comparison, the values of $U^{(\Lambda)}_H$ obtained from SU(3)
models \cite{Weissenborn11b,Tsubakihara10} are also plotted. }
  \end{figure}

Before employing our  EOSs for the hyperonic matter 
to study the properties of the compact stars, we would like 
to look into the behaviour of the $\Lambda-$H potentials
$U^{(\Lambda)}_H$. The values of $\Lambda-$H for given meson-hyperon
coupling strengths are obtained as, 
\begin{equation}
U^{(\Lambda)}_H (\rho) = -g_{\sigma H}\sigma(\rho)+g_{\omega
H}\omega(\rho)+g_{\phi H}\phi(\rho)
\label{eq:ulh}
\end{equation}
where, the values of the fields $\sigma, \omega$ and $\phi$ are
calculated for the pure $\Lambda$ matter at a given density $\rho$. In
Fig.  \ref{fig:ulh}, we plot the results for the  $U^{(\Lambda)}_H$
obtained using the meson-hyperon coupling strengths corresponding
to XSU6 and X180.  For the comparison, we also plot the values of
$U^{(\Lambda)}_H$ obtained for two different models based on the SU(3)
symmetry \cite{Weissenborn11b,Tsubakihara10}. The green dashed curve
is obtained using Ref. \cite{Weissenborn11b} for $Z=0$ which yields
highest value for the maximum mass for the compact star.  The values of
$U^{(\Lambda)}_\Lambda$ and $U^{(\Lambda)}_\Sigma$ for the case of X180
lie within those obtained using other models.  Whereas, the potential 
$U^{(\Lambda)}_\Xi$ is somewhat stiffer in comparison to the other
models as considered.

\section{Compact stars}
\label{sec:cs}

The properties of static or non-rotating compact stars for a given
EOS is obtained by solving the Tolman-Oppenheimer-Volkoff (TOV)
equations \cite{Weinberg72}.  For the case of  rotating compact stars we
solve the Einstein equations for stationary axi-symmetric spacetime.
The numerical computations are performed using the code written by
Stergioulas \cite{Stergioulas95}.
We describe the  outer crust region of the compact star using  the EOS
of R\"uster {\it et~al.} \cite{Ruster06} which  is the recent update of
the  one  given by  Baym, Pethick, and Sutherland \cite{Baym71}. Due
to the fact that the detailed EOS of inner crust indeed is not yet
certain, the polytropic pressure-energy density relation is used to
interpolate the EOS for the region between outer crust and the core
\cite{Fattoyev10}.  The core is assumed to be composed of either the
nucleonic or the hyperonic matter in $\beta-$equilibrium.  The EOS of
the core is obtained from the different parameter sets of the RMF models
as presented in Sec. \ref{sec:eos}.

\begin{figure}
\resizebox{6.0in}{!}{ \includegraphics[]{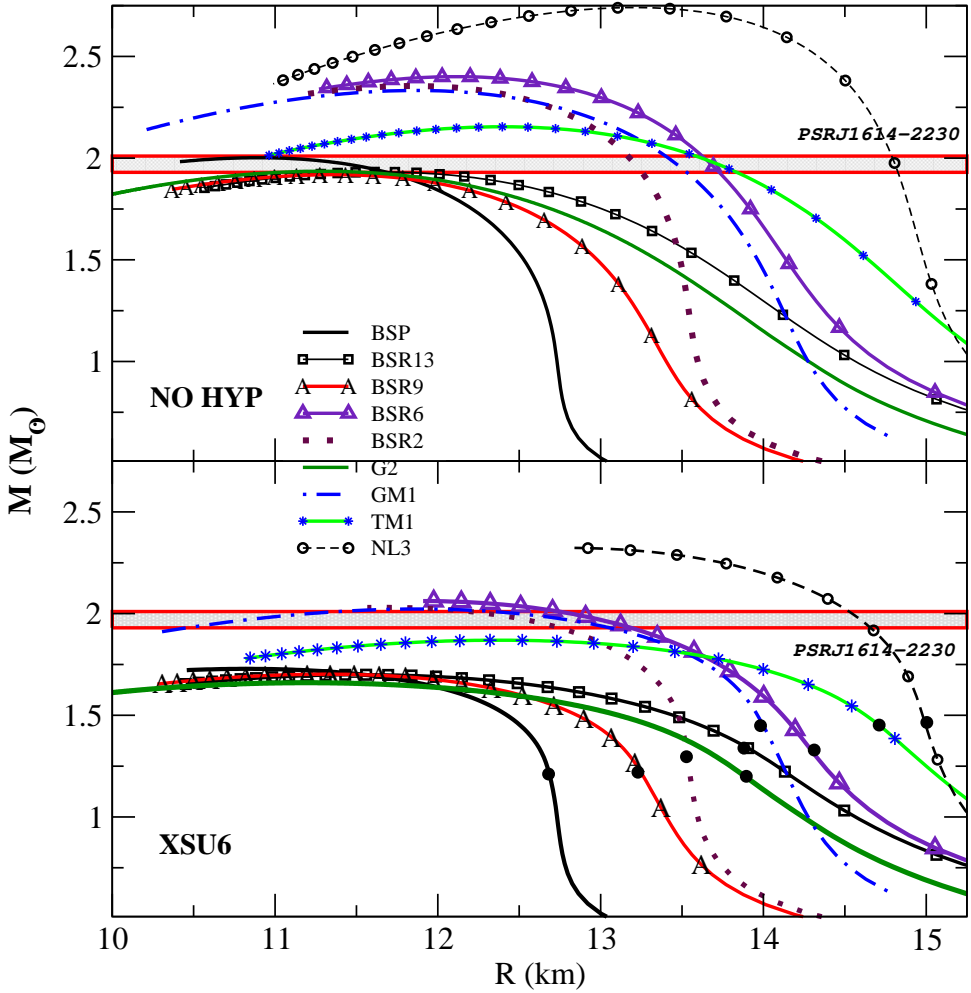}}
\caption{\label{fig:m-r} (Color online)
Plots for the mass-radius relationships for the  equilibrium sequences
of static compact stars obtained using various EOSs for RMF models.
The results with no hyperons are depicted in the upper panel and those
with hyperons in the lower panel. The label XSU6 in the lower panel
represents that the hyperons are included using standard values for the
meson-hyperon coupling strengths.  The curves on the left of the solid
circles represent the equilibrium sequences for the compact stars with
core composed of the hyperonic matter.}
  \end{figure}

In Fig. \ref{fig:m-r} we display our results for the
mass-radius relationships for the static compact stars composed of
$\beta$-equilibrated nucleonic matter (upper panel) and hyperonic matter
(lower panel).  The shaded region represents the current lower limit on
the maximum mass, $M = 1.97\pm 0.04M_\odot$, of the compact star obtained
from the recent mass measurement of the PSRJ1614-2230.  The mass-radius
relationships for the compact stars containing hyperons are obtained using
standard values for the  meson-hyperon coupling strengths as discussed
in the previous section.  It is evident from the upper panel that the
maximum masses of the compact star in the absence of hyperons for all
the RMF models considered are compatible with the current limit on its
lower bound.  The solid circles in the lower panel  indicate the minimum
mass of the compact star at which the hyperons begin to appear. This
minimum mass ranges from $1.2 - 1.5M_\odot$  for the various RMF models
considered.  The parameter sets NL3, GM1, BSR2 and BSR6 yield maximum
mass larger than $2M_\odot$ when the hyperons are included. Thus, these
parameter sets readily allow the hyperons to appear in the PSRJ1614-2230.
The nuclear EOSs, however, for these cases are not compatible with the
heavy-ion data (see Fig. \ref{fig:snm}).  On the other hand, the nuclear
EOSs for the parameter sets BSR9, BSR13, BSP, G2 and TM1 are compatible
with the heavy-ion data, but, they yield the maximum mass in the range
of $1.7 - 1.9 M_\odot$ with the inclusion of hyperons.  It seems that
the nuclear EOSs, compatible with heavy-ion data, can possibly allow the
hyperons to exist in the PSRJ1614-2230 for substantially larger values
of the meson-hyperon coupling strengths in comparison to their standard
values.

\begin{figure}
\resizebox{6.0in}{!}{ \includegraphics[]{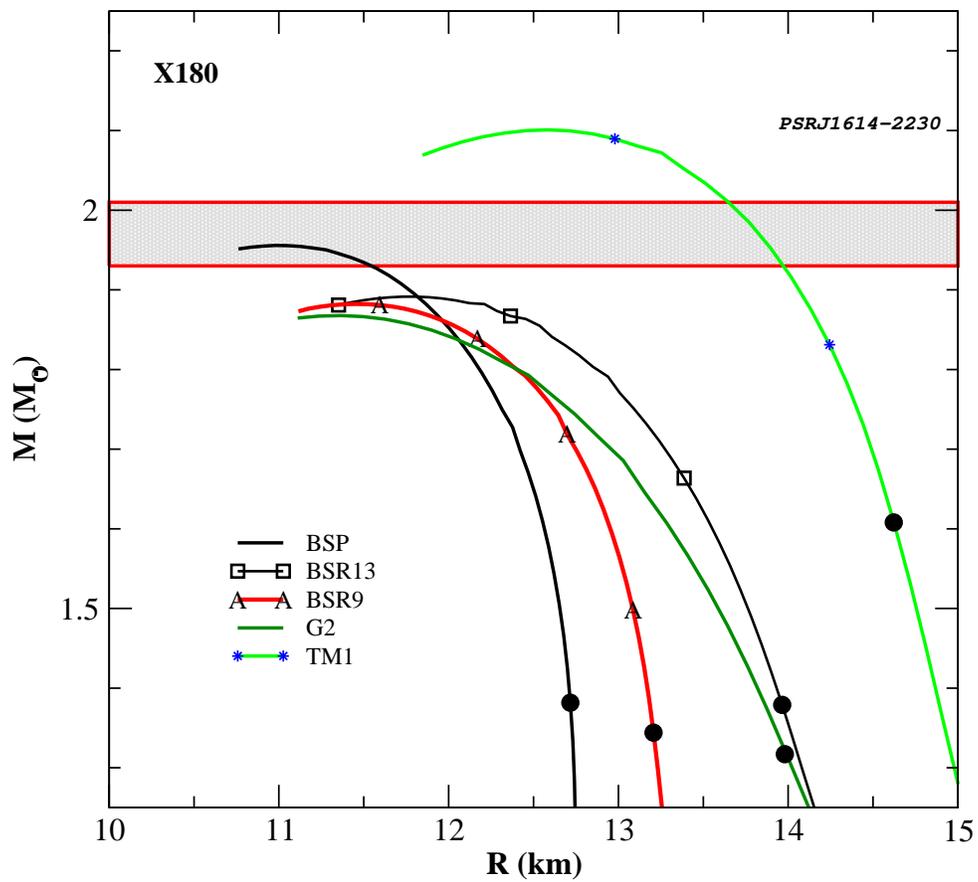}}
\caption{\label{fig:mr_x180} (Color online)
The mass-radius relationships for the static compact  stars obtained
for the values of the meson-hyperon coupling strengths corresponding
to X180.  The curves on the left of the solid circles represent the
sequences for the compact stars with core composed of the
hyperonic matter.  }
  \end{figure}

In Fig. \ref{fig:mr_x180} we show the mass-radius relationships for the
compact stars obtained by augmenting the meson-hyperon coupling strengths
in comparison to their standard values. As discussed previously, the
label  X180 corresponds to $X_{\omega H} = 1$ and $ X_{\phi H} = 0.8$ which is larger
by a factor of $\sim 1.5 $ with respect to their slandered values
taken from the SU(6) quark model.  We plot the results only for those RMF
model for which the nuclear EOSs are compatible with the heavy-ion data.
We find that the maximum masses for the TM1 and BSP parameter sets
are consistent with the mass of the PSRJ1614-2230. 
It may be pointed out to this end that the radius $R_{1.4}$  for the
compact star with canonical mass  of $1.4M_\odot$ for the BSP case is 12.7
km which is consistent with $R_{1.4} = 10.4 - 12.9$ km  as determined
from the recent observations of both transiently accreting and bursting
sources \cite{Steiner12}.  The values of $R_{1.4}$ determined in Ref.
\cite{Steiner12} are independent of assumptions about the composition of
the core.  We also look into the hyperon fractions at the maximum mass.
In Table \ref{tab:ns} we  present our results for the maximum   mass
configurations obtained for two different choices of meson-hyperon
coupling strengths  corresponding to XSU6 and X180. These results are
compared with their nuclear counterparts having no hyperons.  We also
list the values of the partial mass $M_{\rm max}^{\rm HYP}$ which is
composed of the hyperonic matter.  The values of $M_{\rm max}^{\rm HYP}$
are obtained  by  integrating the density profile, corresponding to the
maximum mass configuration, from the central density to the threshold
density at which hyperons begin to appear.  It is interesting to note
that the ratio $M_{\rm max}^{\rm HYP}/M_{\rm max}$ is sizable and almost
the same for both the choices of the meson-hyperon coupling strengths.
The maximum mass obtained using augmented meson-hyperon coupling strengths
are very close to their nuclear counter parts and  yet significant
fraction ($\sim 60\%$) of the mass is composed of the hyperonic matter.
Further, the radius $R_{\rm max}$ at the maximum  mass for the case
with no hyperons and the one obtained  using augmented meson-hyperon
coupling strengths are almost the same.  Thus, it seems possible
to obtain the maximum mass configuration of the compact star,
containing sizable fraction of hyperons, which masquareds its nuclear
counterpart.  The maximum mass of the compact stars with hyperons
nearly same as their nuclear counterpart has also been found in
Ref. \cite{Weissenborn11b,Tsubakihara10}

\begin{figure}
\resizebox{6.0in}{!}{ \includegraphics[]{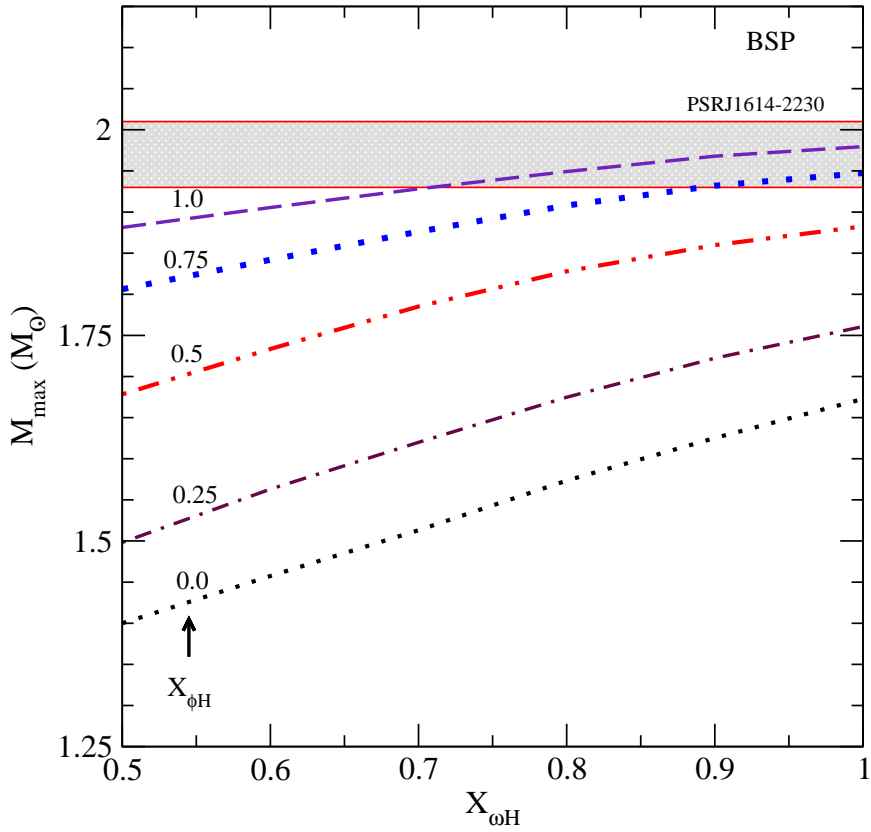}}
\caption{\label{fig:bsr28} (Color online)
The dependence of the maximum mass  $M_{\rm max}$ for the static compact
stars on the values of
meson-hyperon coupling strengths $X_{\omega H}$ and $X_{\phi H}$ (Eqs.
(\ref{eq:xw}) and (\ref{eq:xphi})) for the case of BSP.
}
  \end{figure}

\begin{figure}
\resizebox{6.0in}{!}{ \includegraphics[]{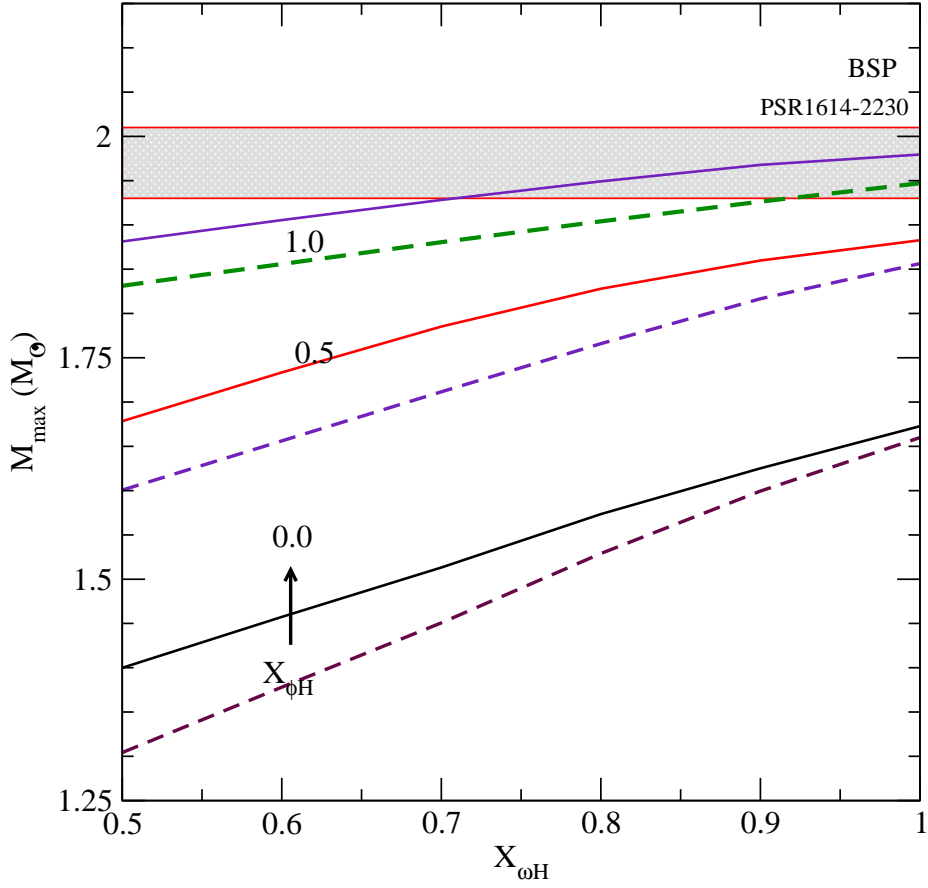}}
\caption{\label{fig:bsr281} (Color online) Similar to
Fig. \ref{fig:bsr28}. But, the values of $M_{\rm max}$ obtained for
$U_{\Sigma}^{(N)}$ = 30 MeV (solid line) are compared with those for
$U_{\Sigma}^{(N)}$ = -30 MeV (dashed line).  }
  \end{figure}

\begin{figure}
\resizebox{6.0in}{!}{ \includegraphics[]{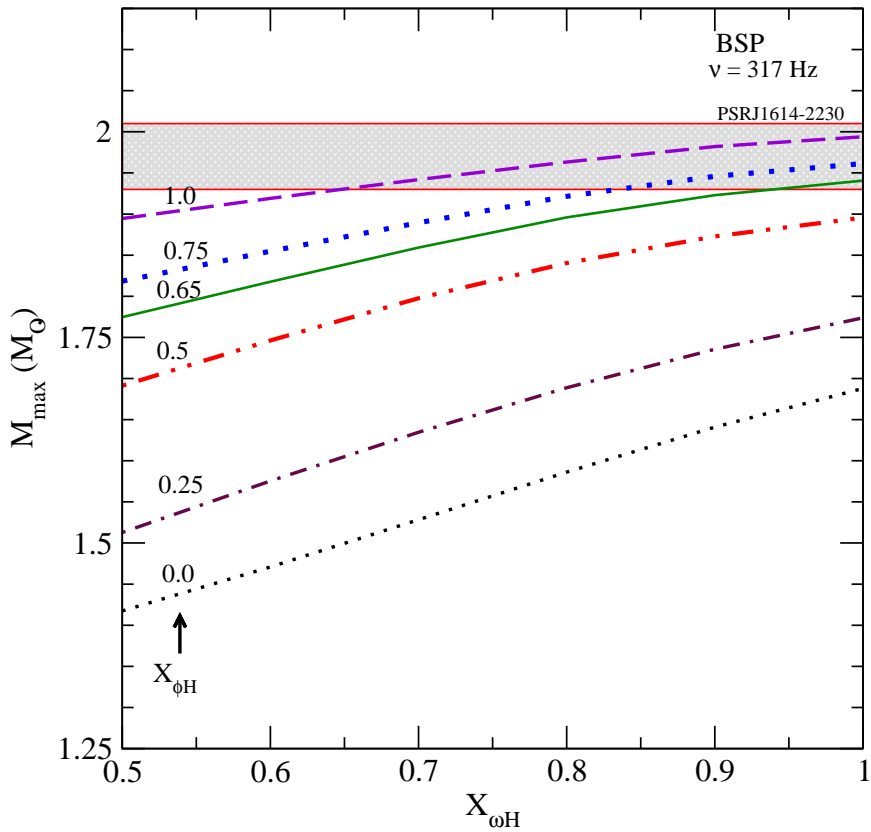}}
\caption{\label{fig:bsp_rns} (Color online) Similar to
Fig. \ref{fig:bsr28}. But, for the compact stars rotating with the
frequency of 317 Hz, same as that of the PSRJ1614-2230.}
  \end{figure}

In Fig. \ref{fig:bsr28} we plot the variations  of $M_{\rm max}$ with
$X_{\omega H}$ and $X_{\phi H}$ for the  case of  BSP.  The value of
$M_{\rm max}$  increases with increasing $X_{\omega H}$ and $X_{\phi H}$.
The $M_{\rm max}$ depends strongly on $X_{\omega H}$ for smaller values
of $X_{\phi H}$. As the $X_{\phi H}$ increases, the value of $M_{\rm max}$
tends to saturate and becomes more or less independent of $X_{\omega H}$.
For $X_{\phi H} = 1$, the $M_{\rm max}$ increases by $0.1M_\odot$ with
increase in  $X_{\omega H}$ from 0.5 to 1.  Also, the $M_{\rm max}$
is less dependent on $X_{\phi H}$ for larger $X_{\omega H}$.  To this
end, we would like to remind that the  $\Sigma - N$ potential depth
$U_\Sigma^{(N)} $ is taken to be 30 MeV. The value of $U_\Sigma^{(N)}$ is
not yet certain. In Fig. \ref{fig:bsr281}, we compare our results for the
$M_{\rm max}$ obtained for $U_\Sigma^{(N)} = -30 $ and 30 MeV.  The value
of $M_{\rm max}$ depends somewhat on $U_\Sigma^{(N)}$ for smaller
values of $X_{\Omega H}$. For $X_{\Omega H} = 0.5$,  the $M_{\rm max}$ may
increases by $\sim 0.1M_\odot$ with increase in $U_\Sigma^{(N)}$ from -30 to
30 MeV.  We also calculate the maximum mass of the compact stars rotating
with the frequency $\nu = 317$ Hz as that of the PSRJ1614-2230. In
Fig. \ref{fig:bsp_rns}, we display the variations  for the $M_{\rm max}$
with $X_{\omega H}$ and $X_{\phi H}$. It can be easily seen that the
value of $X_{\omega H} = 1$ with $X_{\phi H} = 0.65$ is sufficient to
produce the compact star with mass compatible with the current lower
bound on the $M_{\rm max}$.

\section{Conclusions} 
\label{sec:con}

We use various RMF models to explore the possibility of existence of
hyperons in the heaviest observed compact star PSRJ1614-2230 having mass
$M = 1.97 \pm 0.04 M_\odot$.  We have  examined the conditions required
for the hyperons to exist in the PSRJ1614-2230 when  the nuclear EOSs
are subjected to the constrained imposed by the the heavy-ion data.
The various RMF models are selected in such a way that they result in a
diverse set of nuclear EOSs. The values of maximum mass for these models,
in the absence of hyperons, are consistent with the mass of PSRJ1614-2230.
The nuclear EOSs which are consistent with the heavy-ion data require
larger values of the meson-hyperon coupling strengths in order to allow
the hyperons to exist in the PSRJ1614-2230.  Particularly, the coupling
of $\omega$ and $\phi$ vector mesons to the hyperons are required to be
significantly  augmented  with respect to their standard values. These
lager values of the  coupling strengths increases the maximum mass
to the desired limit without significantly affecting the fraction of
maximum mass composed of hyperonic matter.  We find that the maximum
mass configuration of the compact star with sizable fraction of hyperons
masquareds its nuclear counterpart.

\newpage

\begin{thebibliography}{41}
\expandafter\ifx\csname natexlab\endcsname\relax\def\natexlab#1{#1}\fi
\expandafter\ifx\csname bibnamefont\endcsname\relax
  \def\bibnamefont#1{#1}\fi
\expandafter\ifx\csname bibfnamefont\endcsname\relax
  \def\bibfnamefont#1{#1}\fi
\expandafter\ifx\csname citenamefont\endcsname\relax
  \def\citenamefont#1{#1}\fi
\expandafter\ifx\csname url\endcsname\relax
  \def\url#1{\texttt{#1}}\fi
\expandafter\ifx\csname urlprefix\endcsname\relax\def\urlprefix{URL }\fi
\providecommand{\bibinfo}[2]{#2}
\providecommand{\eprint}[2][]{\url{#2}}

\bibitem[{\citenamefont{Demorest et~al.}(2010)\citenamefont{Demorest, Pennucci,
  Ransom, Roberts, and Hessels}}]{Demorest10}
\bibinfo{author}{\bibfnamefont{P.~B.} \bibnamefont{Demorest}},
  \bibinfo{author}{\bibfnamefont{T.}~\bibnamefont{Pennucci}},
  \bibinfo{author}{\bibfnamefont{S.~M.} \bibnamefont{Ransom}},
  \bibinfo{author}{\bibfnamefont{M.~S.~E.} \bibnamefont{Roberts}},
  \bibnamefont{and} \bibinfo{author}{\bibfnamefont{J.~W.~T.}
  \bibnamefont{Hessels}}, \bibinfo{journal}{Nature}
  \textbf{\bibinfo{volume}{467}}, \bibinfo{pages}{1081} (\bibinfo{year}{2010}).

\bibitem[{\citenamefont{Ozel et~al.}(2010)\citenamefont{Ozel, Psaltis, Ransom,
  P.Demorest, and Alford}}]{Ozel10a}
\bibinfo{author}{\bibfnamefont{F.}~\bibnamefont{Ozel}},
  \bibinfo{author}{\bibfnamefont{D.}~\bibnamefont{Psaltis}},
  \bibinfo{author}{\bibfnamefont{S.}~\bibnamefont{Ransom}},
  \bibinfo{author}{\bibnamefont{P.Demorest}}, \bibnamefont{and}
  \bibinfo{author}{\bibfnamefont{M.}~\bibnamefont{Alford}},
  \bibinfo{journal}{Astrophys. J.Lett.} \textbf{\bibinfo{volume}{724}},
  \bibinfo{pages}{L199} (\bibinfo{year}{2010}).

\bibitem[{\citenamefont{Weissenborn et~al.}(2011)\citenamefont{Weissenborn,
  I.Sagret, G.Pagliara, M.Hempel, and Schaffnre-Bielich}}]{Weissenborn11a}
\bibinfo{author}{\bibfnamefont{S.}~\bibnamefont{Weissenborn}},
  \bibinfo{author}{\bibnamefont{I.Sagret}},
  \bibinfo{author}{\bibnamefont{G.Pagliara}},
  \bibinfo{author}{\bibnamefont{M.Hempel}}, \bibnamefont{and}
  \bibinfo{author}{\bibfnamefont{J.}~\bibnamefont{Schaffnre-Bielich}},
  \bibinfo{journal}{Astrophys. J. Lett.} \textbf{\bibinfo{volume}{740}},
  \bibinfo{pages}{L14} (\bibinfo{year}{2011}).

\bibitem[{\citenamefont{Glendenning and Moszkowski}(1991)}]{Glendenning91}
\bibinfo{author}{\bibfnamefont{N.~K.} \bibnamefont{Glendenning}}
  \bibnamefont{and} \bibinfo{author}{\bibfnamefont{S.~A.}
  \bibnamefont{Moszkowski}}, \bibinfo{journal}{Phys. Rev. Lett.}
  \textbf{\bibinfo{volume}{67}}, \bibinfo{pages}{2414} (\bibinfo{year}{1991}).

\bibitem[{\citenamefont{Glendenning and
  Schaffner-Bielich}(1998)}]{Glendenning98}
\bibinfo{author}{\bibfnamefont{N.~K.} \bibnamefont{Glendenning}}
  \bibnamefont{and}
  \bibinfo{author}{\bibfnamefont{J.}~\bibnamefont{Schaffner-Bielich}},
  \bibinfo{journal}{Phys. Rev. Lett.} \textbf{\bibinfo{volume}{81}},
  \bibinfo{pages}{4564} (\bibinfo{year}{1998}).

\bibitem[{\citenamefont{Lackey et~al.}(2006)\citenamefont{Lackey, Nayyar, and
  Owen}}]{Lackey06}
\bibinfo{author}{\bibfnamefont{B.~D.} \bibnamefont{Lackey}},
  \bibinfo{author}{\bibfnamefont{M.}~\bibnamefont{Nayyar}}, \bibnamefont{and}
  \bibinfo{author}{\bibfnamefont{B.}~\bibnamefont{Owen}},
  \bibinfo{journal}{Phys. Rev. D} \textbf{\bibinfo{volume}{73}},
  \bibinfo{pages}{024021} (\bibinfo{year}{2006}).

\bibitem[{\citenamefont{Schulze et~al.}(2006)\citenamefont{Schulze, Polls,
  Ramos, and Vidana}}]{Schulze06}
\bibinfo{author}{\bibfnamefont{H.~J.} \bibnamefont{Schulze}},
  \bibinfo{author}{\bibfnamefont{A.}~\bibnamefont{Polls}},
  \bibinfo{author}{\bibfnamefont{A.}~\bibnamefont{Ramos}}, \bibnamefont{and}
  \bibinfo{author}{\bibfnamefont{I.}~\bibnamefont{Vidana}},
  \bibinfo{journal}{Phys. Rev. C} \textbf{\bibinfo{volume}{73}},
  \bibinfo{pages}{058801} (\bibinfo{year}{2006}).

\bibitem[{\citenamefont{Lattimer and Prakesh}(2007)}]{Lattimer07}
\bibinfo{author}{\bibfnamefont{J.~M.} \bibnamefont{Lattimer}} \bibnamefont{and}
  \bibinfo{author}{\bibfnamefont{M.}~\bibnamefont{Prakesh}},
  \bibinfo{journal}{Phys. Rep.} \textbf{\bibinfo{volume}{442}},
  \bibinfo{pages}{109} (\bibinfo{year}{2007}).

\bibitem[{\citenamefont{Bednarek and Manka}(2007)}]{Bednarek07}
\bibinfo{author}{\bibfnamefont{I.}~\bibnamefont{Bednarek}} \bibnamefont{and}
  \bibinfo{author}{\bibfnamefont{R.}~\bibnamefont{Manka}},
  \bibinfo{journal}{Eur. Phys. Lett.} \textbf{\bibinfo{volume}{78}},
  \bibinfo{pages}{32001} (\bibinfo{year}{2007}).

\bibitem[{\citenamefont{Stone et~al.}(2010)\citenamefont{Stone, Guichon, and
  A.W.Thomas}}]{Stone10}
\bibinfo{author}{\bibfnamefont{J.~R.} \bibnamefont{Stone}},
  \bibinfo{author}{\bibfnamefont{P.}~\bibnamefont{Guichon}}, \bibnamefont{and}
  \bibinfo{author}{\bibnamefont{A.W.Thomas}},
  \bibinfo{journal}{arxiv:1012.2919v1}  (\bibinfo{year}{2010}).

\bibitem[{\citenamefont{Bednarek et~al.}(2012)\citenamefont{Bednarek, Haensel,
  Zdunik, Bejger, and Manka}}]{Bednarek11}
\bibinfo{author}{\bibfnamefont{I.}~\bibnamefont{Bednarek}},
  \bibinfo{author}{\bibfnamefont{P.}~\bibnamefont{Haensel}},
  \bibinfo{author}{\bibfnamefont{J.~L.} \bibnamefont{Zdunik}},
  \bibinfo{author}{\bibfnamefont{M.}~\bibnamefont{Bejger}}, \bibnamefont{and}
  \bibinfo{author}{\bibfnamefont{R.}~\bibnamefont{Manka}},
  \bibinfo{journal}{Astron. Astrophys.} \textbf{\bibinfo{volume}{543}},
  \bibinfo{pages}{A157} (\bibinfo{year}{2012}).

\bibitem[{\citenamefont{Trumper}(2011)}]{Trumper11}
\bibinfo{author}{\bibfnamefont{J.~E.} \bibnamefont{Trumper}},
  \bibinfo{journal}{Prog. Part. Nucl Phys.} \textbf{\bibinfo{volume}{66}},
  \bibinfo{pages}{674} (\bibinfo{year}{2011}).

\bibitem[{\citenamefont{Weissenborn
  et~al.}(2012{\natexlab{a}})\citenamefont{Weissenborn, Chatterjee, and
  Schaffnre-Bielich}}]{Weissenborn11}
\bibinfo{author}{\bibfnamefont{S.}~\bibnamefont{Weissenborn}},
  \bibinfo{author}{\bibfnamefont{D.}~\bibnamefont{Chatterjee}},
  \bibnamefont{and}
  \bibinfo{author}{\bibfnamefont{J.}~\bibnamefont{Schaffnre-Bielich}},
  \bibinfo{journal}{Nucl. Phys. A} \textbf{\bibinfo{volume}{881}},
  \bibinfo{pages}{62} (\bibinfo{year}{2012}{\natexlab{a}}).

\bibitem[{\citenamefont{Logoteta et~al.}(2012)\citenamefont{Logoteta, Bombaci,
  Providencia, and Vidana}}]{Logoteta12}
\bibinfo{author}{\bibfnamefont{D.}~\bibnamefont{Logoteta}},
  \bibinfo{author}{\bibfnamefont{I.}~\bibnamefont{Bombaci}},
  \bibinfo{author}{\bibfnamefont{C.}~\bibnamefont{Providencia}},
  \bibnamefont{and} \bibinfo{author}{\bibfnamefont{I.}~\bibnamefont{Vidana}},
  \bibinfo{journal}{Phys. Rev. D} \textbf{\bibinfo{volume}{85}},
  \bibinfo{pages}{023003} (\bibinfo{year}{2012}).

\bibitem[{\citenamefont{Weissenborn
  et~al.}(2012{\natexlab{b}})\citenamefont{Weissenborn, Chatterjee, and
  Schaffnre-Bielich}}]{Weissenborn11b}
\bibinfo{author}{\bibfnamefont{S.}~\bibnamefont{Weissenborn}},
  \bibinfo{author}{\bibfnamefont{D.}~\bibnamefont{Chatterjee}},
  \bibnamefont{and}
  \bibinfo{author}{\bibfnamefont{J.}~\bibnamefont{Schaffnre-Bielich}},
  \bibinfo{journal}{Phys. Rev. C} \textbf{\bibinfo{volume}{85}},
  \bibinfo{pages}{065802} (\bibinfo{year}{2012}{\natexlab{b}}).

\bibitem[{\citenamefont{Aichelin and Ko}(1985)}]{Aichelin85}
\bibinfo{author}{\bibfnamefont{J.}~\bibnamefont{Aichelin}} \bibnamefont{and}
  \bibinfo{author}{\bibfnamefont{C.~M.} \bibnamefont{Ko}},
  \bibinfo{journal}{Phys. Rev. Lett.} \textbf{\bibinfo{volume}{55}},
  \bibinfo{pages}{2661} (\bibinfo{year}{1985}).

\bibitem[{\citenamefont{Fuchs et~al.}(2001)\citenamefont{Fuchs, Faessler, and
  Zabrodin}}]{Fuchs01}
\bibinfo{author}{\bibfnamefont{C.}~\bibnamefont{Fuchs}},
  \bibinfo{author}{\bibfnamefont{A.}~\bibnamefont{Faessler}}, \bibnamefont{and}
  \bibinfo{author}{\bibfnamefont{E.}~\bibnamefont{Zabrodin}},
  \bibinfo{journal}{Phys. Rev. Lett.} \textbf{\bibinfo{volume}{86}},
  \bibinfo{pages}{1974} (\bibinfo{year}{2001}).

\bibitem[{\citenamefont{Fuchs}(2008)}]{Fuchs08}
\bibinfo{author}{\bibfnamefont{C.}~\bibnamefont{Fuchs}}, \bibinfo{journal}{J.
  Phys. G} \textbf{\bibinfo{volume}{35}}, \bibinfo{pages}{14049}
  (\bibinfo{year}{2008}).

\bibitem[{\citenamefont{Danielewicz et~al.}(2002)\citenamefont{Danielewicz,
  Lynch, and Lacey}}]{Danielewicz02}
\bibinfo{author}{\bibfnamefont{P.}~\bibnamefont{Danielewicz}},
  \bibinfo{author}{\bibfnamefont{W.~G.} \bibnamefont{Lynch}}, \bibnamefont{and}
  \bibinfo{author}{\bibfnamefont{R.}~\bibnamefont{Lacey}},
  \bibinfo{journal}{Science} \textbf{\bibinfo{volume}{298}},
  \bibinfo{pages}{1592} (\bibinfo{year}{2002}).

\bibitem[{\citenamefont{Lamoreaux}(1997)}]{Lamoreaux97}
\bibinfo{author}{\bibfnamefont{S.~K.} \bibnamefont{Lamoreaux}},
  \bibinfo{journal}{Phys. Rev. Lett.} \textbf{\bibinfo{volume}{78}},
  \bibinfo{pages}{5} (\bibinfo{year}{1997}).

\bibitem[{\citenamefont{Decca et~al.}(2005)\citenamefont{Decca, Lo´pez, Chan,
  Fischbach, Krause, and Jamell}}]{Decca05}
\bibinfo{author}{\bibfnamefont{R.~S.} \bibnamefont{Decca}},
  \bibinfo{author}{\bibfnamefont{D.}~\bibnamefont{Lo´pez}},
  \bibinfo{author}{\bibfnamefont{H.~B.} \bibnamefont{Chan}},
  \bibinfo{author}{\bibfnamefont{E.}~\bibnamefont{Fischbach}},
  \bibinfo{author}{\bibfnamefont{D.~E.} \bibnamefont{Krause}},
  \bibnamefont{and} \bibinfo{author}{\bibfnamefont{C.~R.}
  \bibnamefont{Jamell}}, \bibinfo{journal}{Phys. Rev. Lett.}
  \textbf{\bibinfo{volume}{94}}, \bibinfo{pages}{240401}
  (\bibinfo{year}{2005}).

\bibitem[{\citenamefont{Mostepanenko et~al.}(2008)\citenamefont{Mostepanenko,
  Decca, Fischbach, Klimchitskaya, Krause, and L´opez}}]{Mostepanenko08}
\bibinfo{author}{\bibfnamefont{V.~M.} \bibnamefont{Mostepanenko}},
  \bibinfo{author}{\bibfnamefont{R.~S.} \bibnamefont{Decca}},
  \bibinfo{author}{\bibfnamefont{E.}~\bibnamefont{Fischbach}},
  \bibinfo{author}{\bibfnamefont{G.~L.} \bibnamefont{Klimchitskaya}},
  \bibinfo{author}{\bibfnamefont{D.~E.} \bibnamefont{Krause}},
  \bibnamefont{and} \bibinfo{author}{\bibfnamefont{D.}~\bibnamefont{L´opez}},
  \bibinfo{journal}{J. Phys. A} \textbf{\bibinfo{volume}{41}},
  \bibinfo{pages}{164054} (\bibinfo{year}{2008}).

\bibitem[{\citenamefont{Krivoruchenko et~al.}(2009)\citenamefont{Krivoruchenko,
  F.Simkovic, and Faessler}}]{Krivoruchenko09}
\bibinfo{author}{\bibfnamefont{M.~I.} \bibnamefont{Krivoruchenko}},
  \bibinfo{author}{\bibnamefont{F.Simkovic}}, \bibnamefont{and}
  \bibinfo{author}{\bibfnamefont{A.}~\bibnamefont{Faessler}},
  \bibinfo{journal}{Phys. Rev. D} \textbf{\bibinfo{volume}{79}},
  \bibinfo{pages}{125023} (\bibinfo{year}{2009}).

\bibitem[{\citenamefont{Sulaksono et~al.}(2011)\citenamefont{Sulaksono,
  Marliana, and Kasmudin}}]{Sulaksono11}
\bibinfo{author}{\bibfnamefont{A.}~\bibnamefont{Sulaksono}},
  \bibinfo{author}{\bibnamefont{Marliana}}, \bibnamefont{and}
  \bibinfo{author}{\bibnamefont{Kasmudin}}, \bibinfo{journal}{Mod. Phys. Lett.
  A} \textbf{\bibinfo{volume}{26}}, \bibinfo{pages}{367}
  (\bibinfo{year}{2011}).

\bibitem[{\citenamefont{Furnstahl et~al.}(1996)\citenamefont{Furnstahl, Serot,
  and Tang}}]{Furnstahl96}
\bibinfo{author}{\bibfnamefont{R.}~\bibnamefont{Furnstahl}},
  \bibinfo{author}{\bibfnamefont{B.~D.} \bibnamefont{Serot}}, \bibnamefont{and}
  \bibinfo{author}{\bibfnamefont{H.-B.} \bibnamefont{Tang}},
  \bibinfo{journal}{Nucl. Phys.} \textbf{\bibinfo{volume}{A598}},
  \bibinfo{pages}{539} (\bibinfo{year}{1996}).

\bibitem[{\citenamefont{Serot and Walecka}(1997)}]{Serot97}
\bibinfo{author}{\bibfnamefont{B.~D.} \bibnamefont{Serot}} \bibnamefont{and}
  \bibinfo{author}{\bibfnamefont{J.~D.} \bibnamefont{Walecka}},
  \bibinfo{journal}{Int. J. Mod. Phys. E} \textbf{\bibinfo{volume}{6}},
  \bibinfo{pages}{515} (\bibinfo{year}{1997}).

\bibitem[{\citenamefont{Furnstahl et~al.}(1997)\citenamefont{Furnstahl, Serot,
  and Tang}}]{Furnstahl97}
\bibinfo{author}{\bibfnamefont{R.}~\bibnamefont{Furnstahl}},
  \bibinfo{author}{\bibfnamefont{B.~D.} \bibnamefont{Serot}}, \bibnamefont{and}
  \bibinfo{author}{\bibfnamefont{H.-B.} \bibnamefont{Tang}},
  \bibinfo{journal}{Nucl. Phys.} \textbf{\bibinfo{volume}{A615}},
  \bibinfo{pages}{441} (\bibinfo{year}{1997}).

\bibitem[{\citenamefont{Lalazissis et~al.}(1997)\citenamefont{Lalazissis,
  Konig, and Ring}}]{Lalazissis97}
\bibinfo{author}{\bibfnamefont{G.~A.} \bibnamefont{Lalazissis}},
  \bibinfo{author}{\bibfnamefont{J.}~\bibnamefont{Konig}}, \bibnamefont{and}
  \bibinfo{author}{\bibfnamefont{P.}~\bibnamefont{Ring}},
  \bibinfo{journal}{Phys. Rev. C} \textbf{\bibinfo{volume}{55}},
  \bibinfo{pages}{540} (\bibinfo{year}{1997}).

\bibitem[{\citenamefont{Dhiman et~al.}(2007)\citenamefont{Dhiman, Kumar, and
  Agrawal}}]{Dhiman07}
\bibinfo{author}{\bibfnamefont{S.~K.} \bibnamefont{Dhiman}},
  \bibinfo{author}{\bibfnamefont{R.}~\bibnamefont{Kumar}}, \bibnamefont{and}
  \bibinfo{author}{\bibfnamefont{B.~K.} \bibnamefont{Agrawal}},
  \bibinfo{journal}{Phys. Rev. C} \textbf{\bibinfo{volume}{76}},
  \bibinfo{pages}{045801} (\bibinfo{year}{2007}).

\bibitem[{\citenamefont{Agrawal}(2010)}]{Agrawal10}
\bibinfo{author}{\bibfnamefont{B.~K.} \bibnamefont{Agrawal}},
  \bibinfo{journal}{Phys. Rev. C} \textbf{\bibinfo{volume}{81}},
  \bibinfo{pages}{034323} (\bibinfo{year}{2010}).

\bibitem[{\citenamefont{Gal and Millener}(2011)}]{Gal11}
\bibinfo{author}{\bibfnamefont{A.}~\bibnamefont{Gal}} \bibnamefont{and}
  \bibinfo{author}{\bibfnamefont{D.}~\bibnamefont{Millener}},
  \bibinfo{journal}{Phys. Lett.} \textbf{\bibinfo{volume}{B701}},
  \bibinfo{pages}{342} (\bibinfo{year}{2011}).

\bibitem[{\citenamefont{Schaffner-Bielich and Gal}(2000)}]{Schaffner00}
\bibinfo{author}{\bibfnamefont{J.}~\bibnamefont{Schaffner-Bielich}}
  \bibnamefont{and} \bibinfo{author}{\bibfnamefont{A.}~\bibnamefont{Gal}},
  \bibinfo{journal}{Phys. Rev. C} \textbf{\bibinfo{volume}{62}},
  \bibinfo{pages}{034311} (\bibinfo{year}{2000}).

\bibitem[{\citenamefont{Tsubakihara et~al.}(2010)\citenamefont{Tsubakihara,
  Maekawa, Matsumiya, and Ohnishi}}]{Tsubakihara10}
\bibinfo{author}{\bibfnamefont{K.}~\bibnamefont{Tsubakihara}},
  \bibinfo{author}{\bibfnamefont{H.}~\bibnamefont{Maekawa}},
  \bibinfo{author}{\bibfnamefont{H.}~\bibnamefont{Matsumiya}},
  \bibnamefont{and} \bibinfo{author}{\bibfnamefont{A.}~\bibnamefont{Ohnishi}},
  \bibinfo{journal}{Phys. Rev. C} \textbf{\bibinfo{volume}{81}},
  \bibinfo{pages}{065206} (\bibinfo{year}{2010}).

\bibitem[{\citenamefont{Weinberg}(1972)}]{Weinberg72}
\bibinfo{author}{\bibfnamefont{S.}~\bibnamefont{Weinberg}},
  \emph{\bibinfo{title}{Gravitation and Cosmology}} (\bibinfo{publisher}{Wiley,
  New York}, \bibinfo{year}{1972}).

\bibitem[{\citenamefont{Stergioulas and Friedman}(1995)}]{Stergioulas95}
\bibinfo{author}{\bibfnamefont{N.}~\bibnamefont{Stergioulas}} \bibnamefont{and}
  \bibinfo{author}{\bibfnamefont{J.~L.} \bibnamefont{Friedman}},
  \bibinfo{journal}{Astrophys. J.} \textbf{\bibinfo{volume}{444}},
  \bibinfo{pages}{306} (\bibinfo{year}{1995}).

\bibitem[{\citenamefont{Ruster et~al.}(2006)\citenamefont{Ruster, Hampel, and
  Schaffner-Bielich}}]{Ruster06}
\bibinfo{author}{\bibfnamefont{S.~B.} \bibnamefont{Ruster}},
  \bibinfo{author}{\bibfnamefont{M.}~\bibnamefont{Hampel}}, \bibnamefont{and}
  \bibinfo{author}{\bibfnamefont{J.}~\bibnamefont{Schaffner-Bielich}},
  \bibinfo{journal}{Phys. Rev. C} \textbf{\bibinfo{volume}{73}},
  \bibinfo{pages}{035804} (\bibinfo{year}{2006}).

\bibitem[{\citenamefont{Baym et~al.}(1971)\citenamefont{Baym, Pethick, and
  Sutherland}}]{Baym71}
\bibinfo{author}{\bibfnamefont{G.}~\bibnamefont{Baym}},
  \bibinfo{author}{\bibfnamefont{C.}~\bibnamefont{Pethick}}, \bibnamefont{and}
  \bibinfo{author}{\bibfnamefont{P.}~\bibnamefont{Sutherland}},
  \bibinfo{journal}{Astrophys. J.} \textbf{\bibinfo{volume}{170}},
  \bibinfo{pages}{299} (\bibinfo{year}{1971}).

\bibitem[{\citenamefont{F.J.Fattoyev et~al.}(2010)\citenamefont{F.J.Fattoyev,
  C.J.Horowitz, Piekarewicz, and G.Shen}}]{Fattoyev10}
\bibinfo{author}{\bibnamefont{F.J.Fattoyev}},
  \bibinfo{author}{\bibnamefont{C.J.Horowitz}},
  \bibinfo{author}{\bibfnamefont{J.}~\bibnamefont{Piekarewicz}},
  \bibnamefont{and} \bibinfo{author}{\bibnamefont{G.Shen}},
  \bibinfo{journal}{Phys. Rev. C} \textbf{\bibinfo{volume}{82}},
  \bibinfo{pages}{055803} (\bibinfo{year}{2010}).

\bibitem[{\citenamefont{Steiner et~al.}(2012)\citenamefont{Steiner, Lattimer,
  and Brown}}]{Steiner12}
\bibinfo{author}{\bibfnamefont{A.~W.} \bibnamefont{Steiner}},
  \bibinfo{author}{\bibfnamefont{J.~M.} \bibnamefont{Lattimer}},
  \bibnamefont{and} \bibinfo{author}{\bibfnamefont{E.~F.} \bibnamefont{Brown}},
  \bibinfo{journal}{arxiv:1205.26871}  (\bibinfo{year}{2012}).

\bibitem[{\citenamefont{Agrawal et~al.}(2012)\citenamefont{Agrawal, Sulaksono,
  and Reinhard}}]{Agrawal12}
\bibinfo{author}{\bibfnamefont{B.~K.} \bibnamefont{Agrawal}},
  \bibinfo{author}{\bibfnamefont{A.}~\bibnamefont{Sulaksono}},
  \bibnamefont{and} \bibinfo{author}{\bibfnamefont{P.~G.}
  \bibnamefont{Reinhard}}, \bibinfo{journal}{Nucl. Phys. A}
  \textbf{\bibinfo{volume}{882}}, \bibinfo{pages}{1} (\bibinfo{year}{2012}).

\bibitem[{\citenamefont{Y.Sugahara and H.Toki}(1994)}]{Sugahara94}
\bibinfo{author}{\bibnamefont{Y.Sugahara}} \bibnamefont{and}
  \bibinfo{author}{\bibnamefont{H.Toki}}, \bibinfo{journal}{Nucl. Phys.}
  \textbf{\bibinfo{volume}{A579}}, \bibinfo{pages}{557} (\bibinfo{year}{1994}).

\end{thebibliography}

\newpage
\begin{table}
\caption{\label{tab:rmf}
Various  self-interaction and cross-interaction terms included in the
Lagrangian density associated with different parameterizations of the
RMF models considered in the present work. The index '1' and '0' is used
to indicate whether or not the corresponding term is included.  }

 \begin{ruledtabular}
  \begin{tabular}{cccccccc}
\multicolumn{1}{c}{Parameter}&
\multicolumn{3}{c}{Self-interaction}&
\multicolumn{3}{c}{Cross-interaction}&
\multicolumn{1}{c}{Ref.}\\
\cline{2-7}
 & $\sigma$& $\omega$& $\rho$& $\sigma\!-\!\omega$
&
$\sigma\!-\!\rho$
&
\!$\omega\!-\!\rho$\\
BSP& 1& 0& 0&1& 0& 1& \cite{Agrawal12}\\
BSR13& 1& 1& 0&1& 1& 1 & \cite{Dhiman07,Agrawal10}\\
BSR9& 1& 1& 0&1& 1& 1 & \cite{Dhiman07,Agrawal10}\\
BSR6& 1& 0& 0&1& 1& 1 & \cite{Dhiman07,Agrawal10}\\
BSR2& 1& 0& 0&1& 1& 1& \cite{Dhiman07,Agrawal10}\\
G2& 1& 1& 0&1& 1& 0& \cite{Furnstahl97}\\
TM1& 1& 1& 0&0& 0& 0& \cite{Sugahara94}\\
GM1& 1& 0& 0&0& 0& 0&\cite{Glendenning91}\\
NL3& 1& 0& 0&0& 0& 0&\cite{Lalazissis97}\\
 \end{tabular}
 \end{ruledtabular}
 \end{table}

\begin{table}
\caption{\label{tab:nmpro}
Some bulk properties of the nuclear matter at the saturation density
($\rho_s$): binding energy per nucleon ($B/A$), incompressibility
coefficient for symmetric nuclear matter ($K$), symmetry energy ($E_{\rm
sym}(\rho_s)$), linear density dependence of the symmetry energy ($L$)
and various quantities ($K_{\rm sym}$), ($K_{\rm asy}$) and ($K_{\rm
sat2}$) as given by Eqs. (\ref{eq:k0}-\ref{eq:ksat2}).  The values for
the neutron-skin thickness $\Delta R$ for the $^{208}\text{Pb}$ nucleus
are also listed.  The values of $\rho_s$ are in fm$^{-3}$, $\Delta R$
in fm and all the other quantities are in MeV.
 }

  \begin{ruledtabular}
\begin{tabular}{|cccccccccc|}
Parameter& $\Delta _R$& $B/A$&$\rho_s$& $K$& $E_{\rm sym}(\rho_s)$&$ L$&$ K_{\rm sym}$& $K_{\rm asy}$&
$K_{\rm sat2}$\\
\hline
BSP  &  0.15&   15.9&  0.149& 230&   28.83 &  50 &   9& -290& -218\\
BSR13& 0.26 &16.1  &0.147 & 229  & 35.6 &  91&  -40 &-585& -466  \\
BSR9& 0.18  &16.1  &0.147  &233  & 31.6 &  64&  -12 &-396& -313 \\
BSR6& 0.26  &16.1  &0.149 & 236  & 35.4 &  86&  -48 &  -562& -557 \\
BSR2& 0.18  &16.0  &0.149 & 240  & 31.4 &  62&   -4 &  -376& -363 \\ 
G2  &0.26&  16.1   &0.153&  215 &  36.4&  100&   -7& -611  &-404\\
GM1& 0.23 &16.3 & 0.153 &300 &32.5 &94 & 18 &-545&-466\\ 
TM1& 0.27 & 16.3 &   0.145&  281&   36.8 & 111  & 34& -632& -518 \\
NL3 & 0.28 & 16.3 &   0.148&  272 &  37.4&  118&  100& -608& -700 \\
\end{tabular}
\end{ruledtabular}
\end{table}

\begin{table}
\caption{\label{tab:ns}
The values of central density $\rho_c$ (in fm$^{-3}$), radius $R_{\rm max}$
(in km) at the maximum mass $M_{\rm max}$ (in $M_\odot$) for the cases
with no hyperons and with hyperons.  The quantity $M_{\rm max}^{\rm
HYP}$ represents the partial mass  composed of the hyperonic matter.
The label XSU6 denotes the standard values for the meson-hyperon
coupling strengths. The label X180 represents augmented values of the
meson-hyperon coupling strengths.  }

  \begin{ruledtabular}
\begin{tabular}{|cccccccccccccc|}
\multicolumn{1}{|c}{Parameter}&
\multicolumn{3}{c}{NO HYP}&
\multicolumn{5}{c}{XSU6}&
\multicolumn{5}{c|}{X180}\\
\cline{2-14}
& $\rho_c $&$R_{\rm max}$&$ M_{\rm max}$ & $\rho_c $&$R_{\rm max}$&$
M_{\rm max}$&$M_{\rm max}^{\rm HYP}$ &$\frac{M_{\rm max}^{\rm HYP}}{M_{\rm max}}$& $\rho_c $&$R_{\rm max}$&$ M_{\rm
max}$&$M_{\rm max}^{\rm HYP}$&$\frac{M_{\rm max}^{\rm HYP}}{M_{\rm max}}$\\
 \hline
BSP& 1.04& 10.91& 2.00& 1.12& 10.83& 1.73& 1.14& 0.66& 1.08& 10.98& 1.96& 1.29& 0.66\\
BSR13 &0.98&11.66&1.93&1.06&11.53&1.7&1.11&0.65&1.09&11.81&1.89&1.25&0.66\\
BSR9 &1.00&11.36&1.92&1.07&11.26&1.7&1.07&0.63&1.02&11.44&1.88&1.16&0.60\\
BSR6 &0.82&12.12&2.40&0.92&11.91&2.06&1.46&0.71&0.86&12.16&2.34&1.61&0.68\\
BSR2 &0.85&11.84&2.35&0.94&11.65&2.03&1.4&0.69&0.87&11.94&2.29&1.48&0.64\\
G2&1.07& 11.29& 1.93& 1.16& 11.19& 1.66& 1.10& 0.66& 1.08& 11.36& 1.87&
1.23& 0.65\\
GM1 &0.86&11.86&2.33&0.92&11.89&2.02&1.32&0.65&0.88&12.05&2.25&1.37&0.60\\
TM1&0.85&12.39&2.15&0.90&12.38&1.87&1.16&0.62&0.85&12.58&2.10&1.24&0.59\\
NL3
&0.67&13.19&2.74&0.76&12.83&2.32&1.62&0.70&0.69&13.35&2.65&1.73&0.65\\
\end{tabular}
\end{ruledtabular}
\end{table}

\newpage

\end{document}